\documentclass{article}

\usepackage{arxiv}

\usepackage[utf8]{inputenc} 
\usepackage[T1]{fontenc}    
\usepackage{hyperref}       
\usepackage{url}            
\usepackage{booktabs}       
\usepackage{nicefrac}       
\usepackage{microtype}      
\usepackage{lipsum}		
\usepackage{graphicx}
\usepackage{doi}

\usepackage{lineno}
\usepackage{bbm}
\usepackage{amsmath,amsfonts,amssymb,amsthm}

\title{Reconstruction of the temporal correlation network of all-cause mortality fluctuation across Italian regions: the importance of temperature and among-nodes flux}

\author{ \href{https://orcid.org/0000-0003-1899-4579}{\includegraphics[scale=0.06]{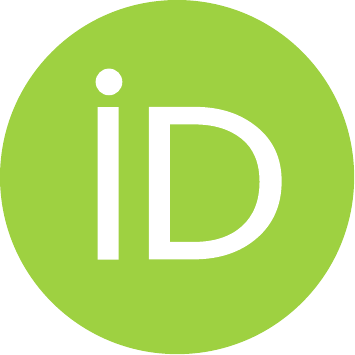}\hspace{1mm}Guido Gigante} \\
	Radiation Protection and Computational Physics\\
	Istituto Superiore di Sanità\\
	00161 Rome, Italy \\
	\texttt{guido.gigante@iss.it} \\
	\And
	\href{https://orcid.org/0000-0002-4640-804X}{\includegraphics[scale=0.06]{orcid.pdf}\hspace{1mm}Alessandro Giuliani} \\
	Environment and Health Department\\
	Istituto Superiore di Sanità\\
	00161 Rome, Italy \\
	\texttt{alessandro.giuliani@iss.it} \\
}

\renewcommand{\shorttitle}{\textit{arXiv} Template}

\hypersetup{
pdftitle={Reconstruction of the temporal correlation network of all-cause mortality fluctuation across Italian regions: the importance of temperature and among-nodes flux},
pdfsubject={q-bio.PE, q-bio.QM},
pdfauthor={Guido Gigante, Alessandro Giuliani},
pdfkeywords={Complex Networks, Dynamical Systems, Epidemiology, Time series},
}

\begin{document}
\maketitle

\begin{abstract}
	All-cause mortality is a very coarse grain, albeit very reliable, index to check the health implications of lifestyle determinants \cite{english2021evaluation}, systemic threats \cite{bilinski2020covid} and socio-demographic factors \cite{foster2018effect}. In this work we adopt a statistical-mechanics approach to the analysis of temporal fluctuations of all-cause mortality, focusing on the correlation structure of this index across different regions of Italy. The correlation network among the 20 Italian regions was reconstructed using temperature oscillations and travellers' flux (as a function of distance and region's attractiveness, based on GDP), allowing for a separation between infective and non-infective death causes. The proposed approach allows monitoring of emerging systemic threats in terms of anomalies of correlation network structure.
\end{abstract}


\section{Introduction}
The monthly-based all-cause death rate fluctuations of the 20 Italian regions are highly correlated in time. This happens even in the absence of recognizable macroscopic parameters like massive epidemics. In this work, we tried and build a phenomenological model of the observed between regions correlations based on the travellers' flux among the network having, as nodes, the regions and, as edges, the mutual travellers' fluxes estimated by a simple exponential model having the distance between regions and Gross Domestic Product (GDP) as major determinants. The above model was complemented by the well-known biphasic effect of temperature on all-cause mortality \cite{martinez2018cold,nielsen2011excess}. The problem can be interpreted as the reconstruction of a network wiring in which the between nodes (regions) edge strength corresponds to the observed temporal correlation of the relative death rates fluctuation in time (Y-network) by the network wiring generated by the combination of between nodes fluxes and temperature effects (X-network).

The strategy of analysis was as follows: the (extremely high) between-region correlation was normalised by what was expected by the observed (well-known) biphasic effect of seasonality. The crude effect of seasonality (when partially out) had as consequence the effect of lowering correlations, but we still have a very high residual correlation asking for a more refined model.

The biphasic effect (high mortality in winter and summer) on all-cause mortality was hypothesised as deriving from an infective component prevailing in winter and a non-infective component prevailing in summer. This interpretation stems from the higher diffusion of viral infections in winter and cardiovascular (often from older people's dehydration) in summer. The winter (infectious) component was modelled using the between regions travellers' flux (exponentially decaying with distance) complemented by the `attractiveness' of each region proportional to its GDP. Thus we generated a `between-region flux network' using a SIR-like model. The summer (non-infectious) model was formalised using a linear function of the month-specific average temperature of each region. This allows us to take into consideration the effect of local heat waves.

A model encompassing the above-sketched elements (X-network) was fitted to the observed death rates, producing the correlation network (Y-network). This minimalistic model was able to reconstruct the death rate oscillation in time and the observed between-region correlation network with high fidelity (corr = 0.993 and corr = 0.841 respectively).

In this work, we demonstrate that weighted edges correlation networks are a very powerful method in epidemiological studies allowing for tracing the dynamics of mortality (morbidity) patterns and potentially discovering anomalies relevant to public health. Taking into account the most general definition of a system as `...a set of interacting units with relationships among them' \cite{miller1976nature} we can safely state that Italy, as for death-rate fluctuations, due to the high temporal correlation among its regions, is a proper system. This allows for the sensible use of second-order statistics (like correlations are) adding unique information content to environmental and epidemiological studies that, in the great majority of cases, rely on the exploitation of a single variable (\textit{e.g.} death rates fluctuations in a given area) in terms of a set of covariates (\textit{e.g.} pollution, age structure, \textit{etc}.).

\section{Materials and Methods}
In the following, we will call $n_{im}$ the number of deaths recorded in region $i{}$ during the $m$-th month of recording; accordingly, we will denote $T_{im}$ the average temperature in the same region during the same month.

\subsection{Bi-phasic effect of temperature}
To account for temperature effects, we assume that $n_{im}$ is Poisson distributed with $\langle n_{im} \rangle = \lambda_{i}(T_{im})$:
\begin{linenomath}
	\begin{equation}
		\lambda_{i}^{\mathrm{biphasic}}(T) = \mathrm{e}^{-a_{c,i} \, T + b_{c,i}} + \mathrm{e}^{a_{h,i} \, T + b_{h,i}} \label{eq.lambda_biphasic} ,
	\end{equation}
\end{linenomath}
where $a_{c,i} > 0$ and $a_{h,i} > 0$ ($c$ and $h$ stand for `cold' and `hot' respectively). This is a convex function with a minimum at:
\begin{linenomath}
	\begin{equation}
		T_{\mathrm{min},i} = \frac{b_{c,i} - b_{h,i} + \log(a_{c,i} / a_{h,i})}{a_{c,i} + a_{h,i}}.
	\end{equation}
\end{linenomath}
The four parameters ($a_{c,i}$, $b_{c,i}$, $a_{h,i}$, and $b_{h,i}$) are fitted, for each region $i$, by maximising the log-likelihood (through the \texttt{scipy.optimize.minimize} function, with TNC method \cite{2020SciPy-NMeth,nash1984newton}):
\begin{linenomath}
	\begin{equation}
		\mathcal{LL} = \sum_{m} n_{im} \, \log \big(\lambda_{i}^{\mathrm{biphasic}}(T_{im}) \big) - \lambda_{i}^{\mathrm{biphasic}}(T_{im}).
	\end{equation}
\end{linenomath}
The result of the fitting is displayed in Figure~\ref{figure2}a.

The temperature $T_{im}$ is computed by associating the main administrative centre of each region with the three closest weather stations for which we have temperature readings. $T_{im}$ is then taken, for each month, as the weighted average of the three stations, with the weight proportional to the inverse of the distance between each station and the main administrative centre.

\subsection{Analysis of commuter flux}
Denote $c_{ij}$ the number of daily commuters from region $j$ to region $i$; and $d_{ij}$\footnote{The inter-region distance $d_{ij}$ we used is the distance between the main administrative centres (``capoluogo'') of each region.} the distance between the same two regions; note that $c_{ij}$, unlike $d_{ij}$, is not symmetric. The continuous line in Figure~\ref{figure3}a is an exponentially decaying function of the distance:
\begin{linenomath}
	\begin{equation}
		c_{ij}^{\mathrm{dist}}(d) = \kappa \, \big( \mathrm{e}^{-\frac{d}{d_{0}}} + \mathbbm{b}_{0}  \big)
	\end{equation}
\end{linenomath}
where the parameters ($\kappa$, $d_{0}$, and $\mathbbm{b}_{0}$) are fitted by maximising the log-likelihood (through the \texttt{scipy.optimize.minimize} function, with TNC method; the zeros -- no commuters from region $j$ to region $i$ -- are not included in the fit):
\begin{linenomath}
	\begin{equation}
		\mathcal{LL} = -\sum_{i,\,j \mid c_{ij} > 0} \Big( \log \big( c_{ij} \big) - \log \big( c_{ij}^{\mathrm{dist}}(d_{ij}) \big) \Big)^2 .
	\end{equation}
\end{linenomath}
Here we are assuming a log-normal distribution for $c_{ij}$ around the expected value $c_{ij}^{\mathrm{dist}}$.

Defining $c_{i:} \equiv \sum_{j} c_{ij}$ -- the number of commuters to region $i$, and calling $\mathrm{GDP}_{i}$ the GDP of region $i$, in Figure~\ref{figure3}b the continuous line is the fit of the linear function:
\begin{linenomath}
	\begin{equation}
		c_{i:}^{\mathrm{gdp}}(\mathrm{GDP}) = \kappa \, \mathrm{GDP}
	\end{equation}
\end{linenomath}
whose slope $\kappa$ is fitted by maximising the log-likelihood (through the \texttt{scipy.optimize.minimize} function, with TNC method):
\begin{linenomath}
	\begin{equation}
		\mathcal{LL} = -\sum_{i} \Big( \log \big( c_{i:} \big) - \log \big( c_{i:}^{\mathrm{gdp}}(\mathrm{GDP}_{i}) \big) \Big)^2 .
	\end{equation}
\end{linenomath}
We are assuming that $c_{i:}$ follows a log-normal distribution around the expected value $c_{i:}^{\mathrm{gdp}}$. This is in contrast with the assumption of log-normality for $c_{ij}$, since the sum of log-normal variables is not itself log-normal. Yet, in many cases, this is a good approximation \cite{mitchell1968permanence}.

Finally, we performed a fit that considers the two effects together: the exponential decay with distance, and the linear dependence on the GDP of the region of destination; indicating with $\mathrm{pop}_{j}$ the population of region $j$, we have:
\begin{linenomath}
	\begin{equation}
		c_{ij}^{\mathrm{fit}} = \kappa \, \mathrm{pop}_{j} \, \mathrm{GDP}_{i} \, \mathrm{e}^{-\frac{d_{ij}}{d_{0}}} \label{eq.c_dist_gdp} .
	\end{equation}
\end{linenomath}
As above, the fit procedure finds the best parameters $\kappa$ and $d_{0}$ by maximising the log-normal log-likelihood (through the \texttt{scipy.optimize.minimize} function, with TNC method; the zeros -- no commuters from region $j$ to region $i$ -- are not included in the fit). The result is shown in Figure~\ref{figure4}.

\subsection{Total flux}
We hypothesise that the total flux of persons $f_{ij}$ comprises, beyond the daily commuters $c_{ij}$, an ``episodic'' component $e_{ij}$ of more irregular movements: 
\begin{linenomath}
	\begin{equation}
		f_{ij} = c_{ij} + e_{ij} \label{eq.flux_c_e}.
	\end{equation}
\end{linenomath}
Starting from the results for $c_{ij}$, we make the assumption that $e_{ij}$ is an exponentially decaying function of the distance between regions, and a linear function of the GDP of the region of arrival $i{}$ and of the population of the region of departure:
\begin{linenomath}
	\begin{equation}
		e_{ij} = \kappa^{e} \, \mathrm{pop}_{j} \, \mathrm{GDP}_{i} \, \mathrm{e}^{-d_{ij} / d_{0}^{e}} \label{eq.episodic}.
	\end{equation}
\end{linenomath}
With respect to Equation~\ref{eq.c_dist_gdp}, we expect $d_{0}^{e} > d_{0}$, since episodic travels, in contrast with frequent ones, are likely less affected by the distance to travel.

\subsection{SIR network model}
The full flux-temperature model includes two different effects. The first one is related to the non-infective component of mortality:
\begin{linenomath}
	\begin{equation}
		\lambda_{im}^{\mathrm{flux}} = \mathrm{pop}_{i} \, \Big( \mathrm{e}^{a_{h} \, T_{im} + b_{h}} + \rho_{0i} \Big) + \dots \label{eq.lambda_flux_partial} ,
	\end{equation}
\end{linenomath}
where $\lambda_{im}^{\mathrm{flux}}$ is the model expectation for the number of deaths in region $i$ at month $m$; and $\rho_{0i}$ is a baseline mortality rate for region $i$. Note that this effect is akin to the warm-season component of Equation~\ref{eq.lambda_biphasic}, but here in the flux-temperature model, for the sake of parsimony, we lose the individualised behaviour of each region $i$, and all regions respond to high temperatures in the same way.

The second effect takes into account the infective component of mortality. We make the simplifying assumption that, in each month, a new infectious disease starts spreading; at the end of the month, a fraction $\mu$ of the people ``recovered'' from the disease dies; the following month, the process starts afresh. The spreading of the disease follows a SIR (Susceptible, Infected, Recovered) model \cite{hethcote2000mathematics} on the flux network. Defining the two matrices:
\begin{linenomath}
	\begin{align}
		\phi_{ij}                                    & =                                                                                
		\begin{cases}
		\frac{f_{ij}}{\mathrm{pop}_{j}}              & \text{for } i \neq j                                                             \\    
		1 - \frac{\sum_{k} f_{kj}}{\mathrm{pop}_{j}} & \text{for } i = j                                                                
		\end{cases}
		\label{eq.phi} \\
		\hat{\phi}_{ij}                              & = \Big(\frac{\phi_{ij}}{\sum_{k} \phi_{ik}} \Big)^{\intercal} \label{eq.phi_hat} 
	\end{align}
\end{linenomath}
the dynamics of the model reads:
\begin{linenomath}
	\begin{align}
		\dot{S}_{i} & = -\sum_{l} \hat{\phi}_{il} \, \biggl[ \frac{\beta}{\mathrm{pop}_{l}} \, \Big( \sum_{j} \phi_{lj} \, S_{j} \Big) \, \Big( \sum_{j} \phi_{lj} \, I_{j} \Big) \biggl] \label{eq.S}                  \\
		\dot{I}_{i} & = \sum_{l} \hat{\phi}_{il} \, \biggl[ \frac{\beta}{\mathrm{pop}_{l}} \, \Big( \sum_{j} \phi_{lj} \, S_{j} \Big) \, \Big( \sum_{j} \phi_{lj} \, I_{j} \Big) \biggl] - \gamma \, I_{i} \label{eq.I} \\
		\dot{R}_{i} & = \gamma \, I_{i} \label{eq.R} ,                                                                                                                                                                  
	\end{align}
\end{linenomath}
where $S_{i}$, $I_{i}$, and $R_{i}$ are the number, respectively, of susceptible, infected, and recovered individuals in region $i$; $\beta$ measures the rate at which susceptible individuals get infected ($S \rightarrow I$); and $\gamma$ is the rate of recovery, $I \rightarrow R$. The model, therefore, consists of 60 coupled differential equations.

The reasoning behind the model is as follows. The term $\sum_{j} \phi_{ij} \, S_{j}$ represents the number of susceptible individuals, in region $i$, at a given instant of time, due to the flux from other regions (minus the flux out of region $i$ itself -- the diagonal elements $\phi_{ii}$); for the infected, it is $\sum_{j} \phi_{ij} \, I_{j}$. In a classical SIR model, the number of newly infected individuals $\mathrm{d}I$ is given by $\beta \, \frac{S \, I}{\mathrm{pop}}$; in our case:
\begin{linenomath}
	\begin{equation}
		\mathrm{d}I_{l} = \frac{\beta}{\mathrm{pop}_{l}} \, \Big( \sum_{j} \phi_{lj} \, S_{j} \Big) \, \Big( \sum_{j} \phi_{lj} \, I_{j} \Big) \label{eq.dI} .
	\end{equation}
\end{linenomath}
At end of the day, the reverse flux $f_{ji}$ (people moving back from region $i$ to region $j$) will redistribute the newly infected, in proportion to the fraction of susceptible individuals contributed by each region $j$; this is given by:
\begin{linenomath}
	\begin{equation}
		\mathrm{d}I_{i} = \sum_{l} \hat{\phi}_{il} \mathrm{d}I_{l} ,
	\end{equation}
\end{linenomath}
that, together with Equation~\ref{eq.dI}, gives the infinitesimal increment of infected people entering Equation~\ref{eq.I} (first term on the left).

The initial conditions are always in the form of one ``patient zero'' in region $i_{0}$ at time $t = 0$, so that $S^{i}(t=0) = \mathrm{pop}_{i}$ for $i \neq i_{0}$, and $S^{i_{0}}(t=0) = \mathrm{pop}_{i_{0}} - 1$; accordingly, $I^{i}(t=0) = 0$ for $i \neq i_{0}$, and $I^{i_{0}}(t=0) = 1$. Since $i_{0}$ is not known, we assume it to be a random variable, distributed such that:
\begin{linenomath}
	\begin{equation}
		p(i_{0} = i) \propto \mathrm{GDP}_{i} \label{eq.pi0} ;
	\end{equation}
\end{linenomath}
this amounts to assuming that the external flux to region $i$ (people coming to region $i$ from outside Italy) is proportional to the GDP of the region itself.

For each month we evolve Equations~\ref{eq.S}, \ref{eq.I}, and \ref{eq.R} for 30 days ($t \in [0, 30]$); equations are integrated using Euler method, with step size $dt = 1\, \mathrm{day}$. Finally, this infective component of mortality is incorporated in the model:
\begin{linenomath}
	\begin{equation}
		\lambda_{imi_{0}}^{\mathrm{flux}} = \mathrm{pop}_{i} \, \Big( \mathrm{e}^{a_{h} \, T_{im} + b_{h}} + \rho_{0i} \Big) +  \mu \, R_{im}^{i_{0}}(t=30) \label{eq.lambda_flux},
	\end{equation}
\end{linenomath}
where with $R_{im}^{i_{0}}(t=30)$ we designate the total number of recovered individuals for region $i$ at the end of month $m$, when the patient zero was located in region $i_{0}$ (all months are assumed, for simplicity, to have 30 days).

To incorporate seasonal effects also in the infective dynamics we make $\beta$ a function of the temperature:
\begin{linenomath}
	\begin{equation}
		\beta_{im} = \mathrm{e}^{-a_{\beta} \, T_{im} + b_{\beta}} \label{eq.beta_T},
	\end{equation}
\end{linenomath}
with $a_{\beta} > 0$; with the additional constraint that $\beta < \frac{1}{dt}$\footnote{The condition $\beta = \frac{1}{dt}$ amounts to having all the population infected in a single $dt$; larger values lead, in the Euler approximation, to unphysical solutions.}.

Considering Equations~\ref{eq.flux_c_e} and \ref{eq.lambda_flux_partial} (parameters $a_{h}$, $b_{h}$, and $\rho_{0i}$), Equation~\ref{eq.episodic} ($\kappa^{e}$ and $d_{0}^{e}$), Equation~\ref{eq.lambda_flux} ($\mu$), Equation~\ref{eq.beta_T} ($a_{\beta}$ and $b_{\beta}$), alongside Equations~\ref{eq.S}, \ref{eq.I}, and \ref{eq.R} ($\gamma$), the model comprises 28 parameters; of which, 20 (the $\rho_{0i}$) are simply used to offset the different mortality rates in different regions (due, for example, to distinct age structures). These parameters are fitted to the data by maximising the Poisson log-likelihood:
\begin{linenomath}
	\begin{equation}
		\mathcal{LL} = \frac{1}{N_{\mathrm{counts}}} \, \sum_{i_{0}} p(i_{0}) \, \Big[ \sum_{i,\,m} n_{im} \, \log \big(\lambda_{imi_{0}}^{\mathrm{flux}} \big) - \lambda_{imi_{0}}^{\mathrm{flux}} \Big] \label{eq.log_likelihood_flux} ,
	\end{equation}
\end{linenomath}
where $N_{\mathrm{counts}}$ is the number of terms in the sum $\sum_{i, \, m}$\footnote{If we consider $n_{\mathrm{batch}}$ different months, having 20 regions, $N_{\mathrm{counts}} = 20 \cdot n_{\mathrm{batch}}$.}, and $p(i_{0})$ is given by Equation~\ref{eq.pi0}.

To this likelihood, we added two prior likelihoods to constrain the parameters of the model. The first is a (soft\footnote{The log-prior becomes quadratic outside the allowed (flat probability) range; the factor in front of the quadratic term is chosen large enough to practically prevent leaving the allowed range.}) flat prior, not-null in the range $[0, \, 0.2]$ for the quantity $\frac{\sum_{k} f_{ki}}{\mathrm{pop}_{i}}$ (see Equation~\ref{eq.flux_c_e}). This constrains the fraction of the population leaving a region every day to less than $20\%$. The second log-prior is quadratic in $d_{0}^{e}$ (Equation~\ref{eq.episodic}):
\begin{linenomath}
	\begin{equation}
		\log \big( p(d_{0}^{e}) \big) = -9.74 \cdot 10^{-7} \, \big( d_{0}^{e} \big)^2 + \mathrm{const} ,
	\end{equation}
\end{linenomath}
to penalize very high spatial decay constants $d_{0}^{e}$ for the episodic component of the flux.

The maximisation has been carried out, in this case, through the Adam optimiser \cite{kingma2014adam}, with default parameters ($\beta_{1} = 0.9$ and $\beta_{2} = 0.99$) and a learning rate decreasing at each optimisation step according to:
\begin{linenomath}
	\begin{equation}
		lr(\mathrm{step}) = \frac{10^{-3}}{(1 + \frac{\mathrm{step}}{10^4})^{0.75}} .
	\end{equation}
\end{linenomath}

The training set consists of the monthly death counts for each of the 20 regions for 96 (out of 108) randomly chosen months in the period 2011-2019. We reserved 12 months (12 + 96 = 108) as test data; these months were selected to have one exemplar of each calendar month - January to December; since the data set spans only 9 years, 3 randomly-selected years contributed two months (6 months apart, \textit{e.g.} April-October) to the test data.

At each step, $n_{\mathrm{batch}}$ months\footnote{With month here we denote one specific month in a specific year; so, in the training set, we have 96 months.} are randomly selected from the training set (the same month can appear multiple times in the batch). For each month $m$, a patient zero-region $i_{0m}$ is randomly extracted with probability given by Equation~\ref{eq.pi0}. The computed log-likelihood is then:
\begin{linenomath}
	\begin{equation}
		\mathcal{LL}_{\mathrm{batch}} = \frac{1}{20 \cdot n_{\mathrm{batch}}} \, \sum_{i,\,m, \, i_{0,m}} n_{im} \, \log \big(\lambda_{imi_{0m}}^{\mathrm{flux}} \big) - \lambda_{imi_{0m}}^{\mathrm{flux}}  \label{eq.log_likelihood_flux_batch} ,
	\end{equation}
\end{linenomath}
a stochastic approximation of the total log-likelihood of Equation~\ref{eq.log_likelihood_flux}.

For the first $10^4$ optimization steps, $n_{\mathrm{batch}} = 10$. From that step onward, $n_{\mathrm{batch}} = 100$; and, to the log-likelihood, we added a ``regularization'' term:
\begin{linenomath}
	\begin{equation}
		\mathcal{LL}_{\mathrm{corr}} = -\frac{\upsilon_{\mathrm{corr}}}{190} \, \sum_{i > j} \Big( \mathrm{corr}_{ij} - \mathrm{corr}^{0}_{ij} \Big)^2
	\end{equation}
\end{linenomath}
where $\mathrm{corr}^{0}_{ij}$ is the actual correlation between the monthly deaths of region $i$ and region $j$; whereas $\mathrm{corr}_{ij}$ are the corresponding correlations produced by the model (on the specific batch); the factor $\frac{1}{190}$ normalises the sum $\sum_{i>j}$, which comprises 190 terms. We set $\upsilon_{\mathrm{corr}} = 8.76 \cdot 10^{2}$.

We monitored the log-likelihood (Equation~\ref{eq.log_likelihood_flux}) on test data during the training; since it never substantially decreased (that would suggest some level of over-fitting), we interrupted the optimisation after $10^{6}$ steps, when improvement on the training set appeared extremely slow.

All the computations were performed with custom code written in Python; core functions were just-in-time compiled, and their gradient was computed where necessary, through the Jax package (\url{https://github.com/google/jax}).

\section{Results}
\begin{figure}[h!]
	\setlength{\unitlength}{\textwidth}
	\begin{picture}(1,0.45)
		\put(0.,0.05)
		{
			\includegraphics[width=0.45\unitlength]{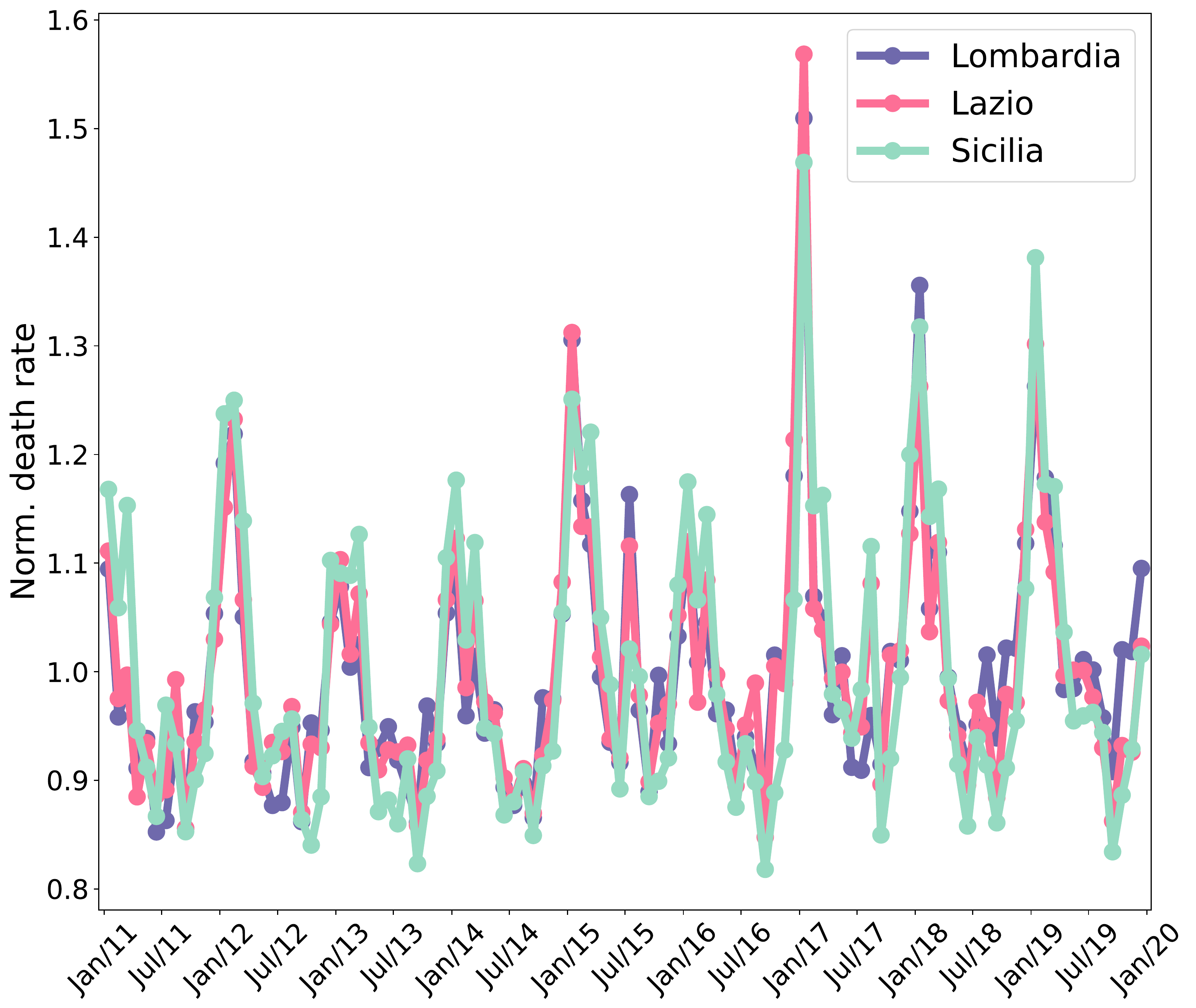}
		}
					   
		\put(0.5,0.05)
		{
			\includegraphics[width=0.45\unitlength]{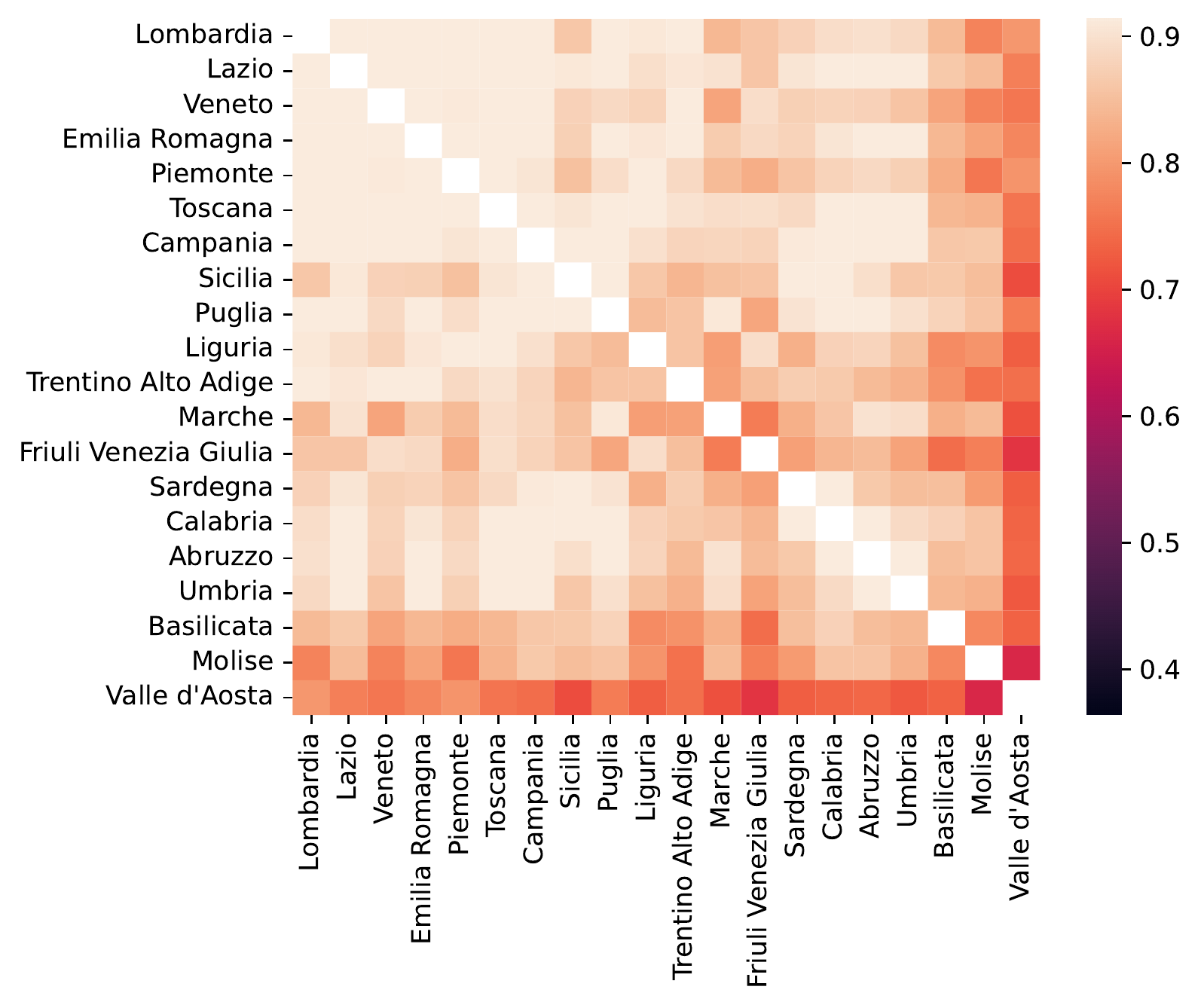}
		}
					   
		\put(0.25,0.){\makebox(0,0)[b]{(a)}}
		\put(0.75,0.){\makebox(0,0)[b]{(b)}}
					   
	\end{picture}
	\caption{Death rates in different regions are extremely correlated. (\textbf{a}) Actual time series for three regions, normalized to have an average value equal to one. The three lines present a strikingly similar course. (\textbf{b}) Pairwise between-region correlations (regions are ordered -- left to right and top to bottom -- according to decreasing GDP). A trend with GDP is appreciable, with smaller and less densely populated regions (\textit{i.e.} Valle d'Aosta and Molise) endowed with lower (albeit, still high) correlations.}
	\label{figure1}
\end{figure}

The course of monthly death rates, normalized to the mean over the entire period, is strikingly similar for different regions. This can be appreciated in Figure~\ref{figure1}a, where we show three regions (chosen to be representative of the north -- Lombardia, centre -- Lazio, and south -- Sicilia, of Italy). 

Such observation is made more quantitative in Figure~\ref{figure1}b, which shows the between-region correlation matrix of monthly death rate fluctuations (correlations computed on 108 data points) relative to the different regions. As evident from the figure, the between-region correlations are extremely high (0.865 ± 0.063), with smaller and less densely populated regions (\textit{i.e.} Valle d'Aosta and Molise) endowed (as expected) by a lower (albeit very significant) average correlation strength (0.739 and 0.805 respectively).

\unskip

To check if the bi-phasic effect of temperature was sufficient to get rid of the observed correlations (in presence of a substantially similar age structure across the different Italian regions), a quantitative model taking into account the temperature effect was fitted to the different regions' mortality data (see Materials and Methods).

\begin{figure}[h!]
	\setlength{\unitlength}{\textwidth}
	\begin{picture}(1,0.45)
		\put(0.,0.05)
		{
			\includegraphics[width=0.45\unitlength]{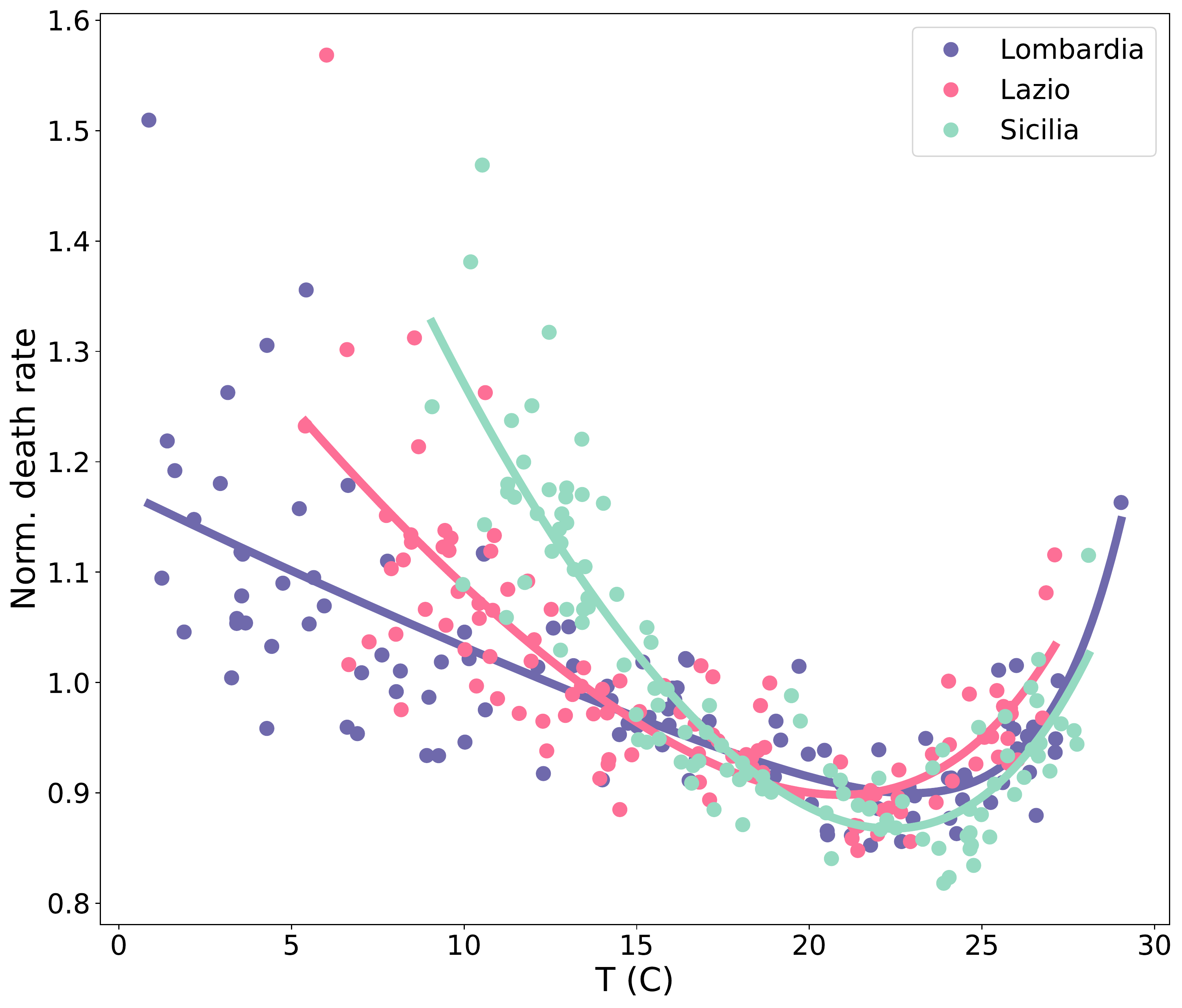}
		}
					   
		\put(0.5,0.05)
		{
			\includegraphics[width=0.45\unitlength]{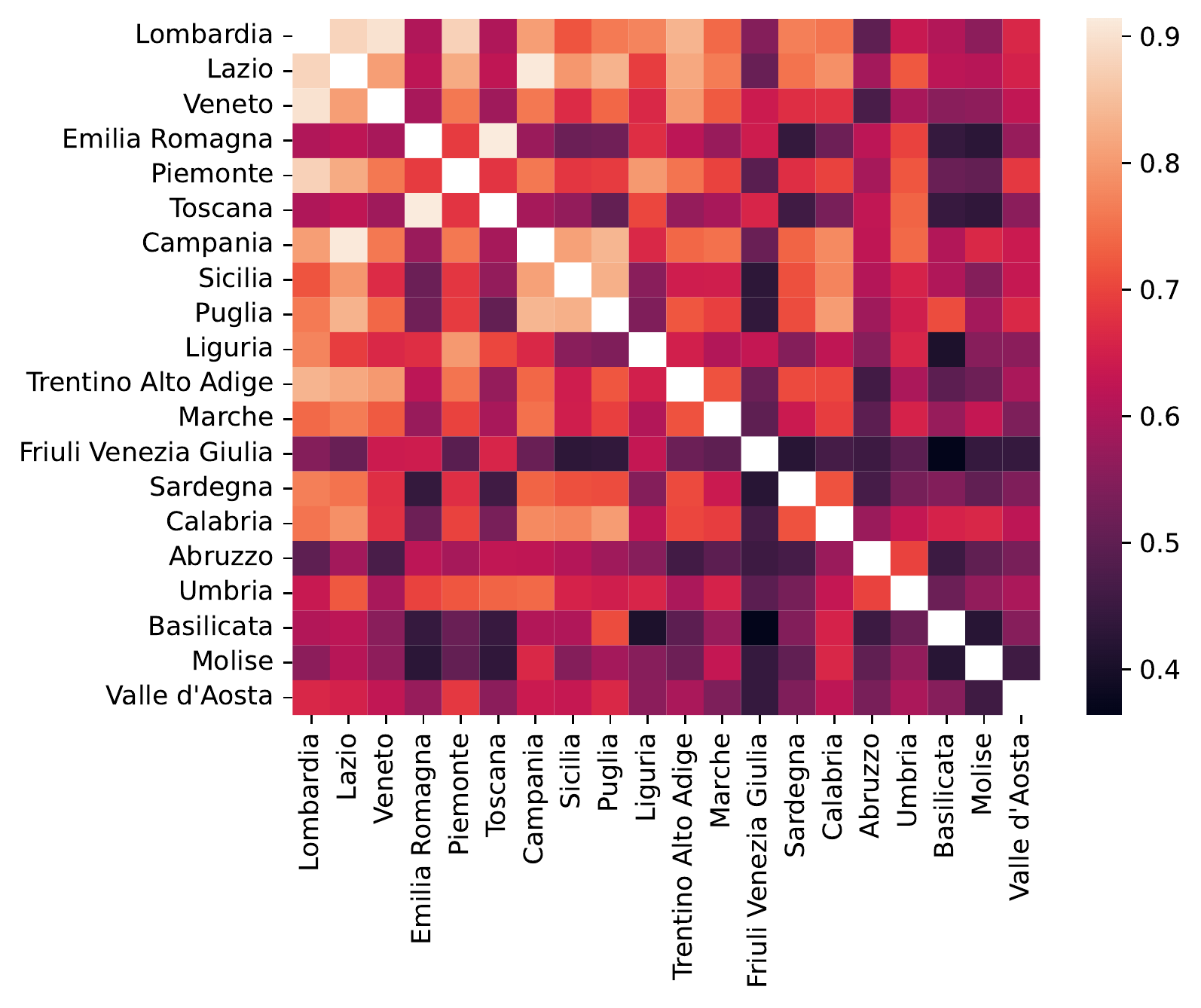}
		}
					   
		\put(0.25,0.){\makebox(0,0)[b]{(a)}}
		\put(0.75,0.){\makebox(0,0)[b]{(b)}}
					   
	\end{picture}
	\caption{(\textbf{a}) Normalized death rates for three regions as a function of the temperature. The continuous lines are the results of a fit (see Materials and Methods). (\textbf{b}) Pairwise between-region correlations for the time series of the deaths, when the fitted effect of the temperature is subtracted from the raw numbers. Correlations drastically decrease (colour scale as in Figure~\ref{figure1}b), but remain large.}
	\label{figure2}
\end{figure}

All regions showed very similar relations between death rate fluctuations and temperature with the expected bi-phasic relation with two winter and summer peaks and a minimum at intermediate temperature values (spring and autumn) (see Figure~\ref{figure2}a reporting three representative regions data).

\unskip

By normalizing the time series of death rate fluctuations by temperature effect, the between-region correlation drastically decreases (0.63 ± 0.12), so confirming the expected effect of temperature on mortality (Figure~\ref{figure2}b; colour scale as in Figure~\ref{figure1}b). Notwithstanding that, the entity of residual correlation is still high, asking for some other relevant factor to be taken into consideration.

\begin{figure}[h!]
	\setlength{\unitlength}{\textwidth}
	\begin{picture}(1,0.41)
		\put(0.,0.05)
		{
			\includegraphics[width=0.45\unitlength]{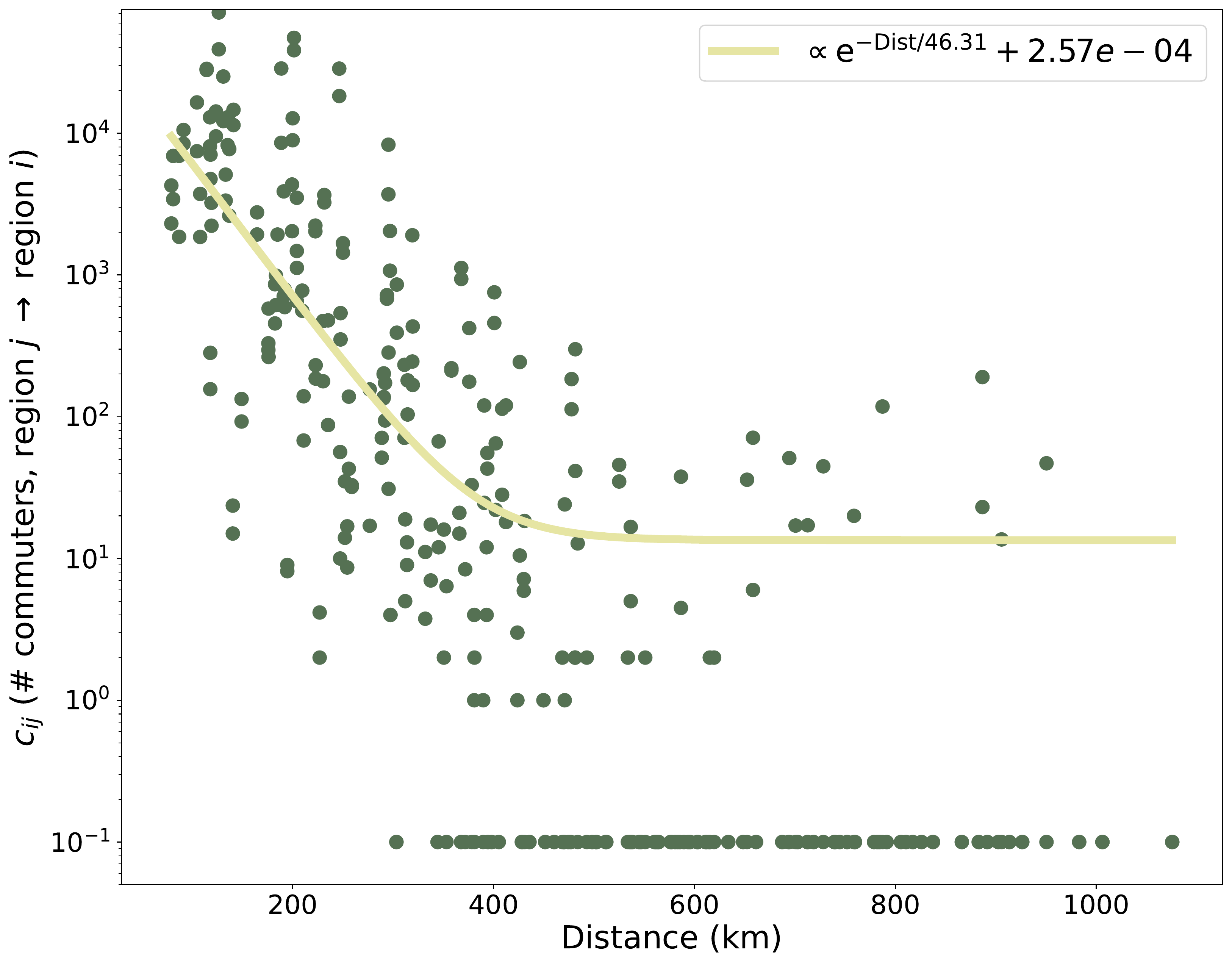}
		}
					   
		\put(0.5,0.05)
		{
			\includegraphics[width=0.45\unitlength]{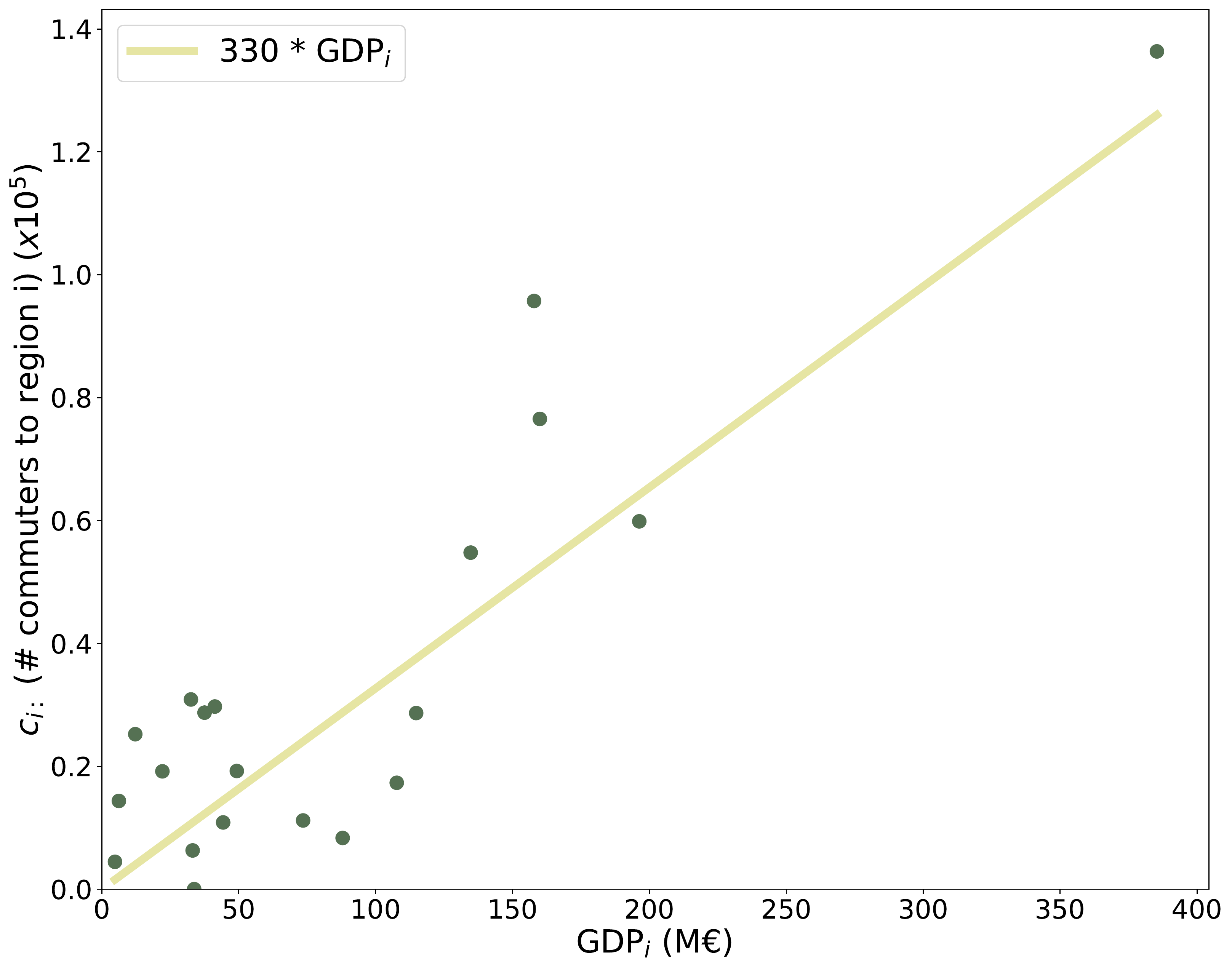}
		}
					   
		\put(0.25,0.){\makebox(0,0)[b]{(a)}}
		\put(0.75,0.){\makebox(0,0)[b]{(b)}}
					   
	\end{picture}
	\caption{Determinants of the commuters' flux. (\textbf{a}) The flux between two regions decays exponentially with the distance to travel (continuous line, exponential fit). The points at the bottom of the graph are zeros (not allowed in logarithmic scale and not considered in the fit). (\textbf{b}) The flux increases linearly with the GDP of the region of destination (continuous line, linear fit).}
	\label{figure3}
\end{figure}

We hypothesise that strong correlations among regions also arise for an infective component, continually spreading from region to region at small time scales (less than a month), driven by the movement of people from one region to another. Firstly we examined the data about the flux of daily commuters between regions; such data shows a clear dependence both on spatial distance (exponential decay; see Figure~\ref{figure3}a; the continuous line is the result of a fit) and the GDP of the region of arrival, having the role of an `attractiveness' factor (linear dependence; see Figure~\ref{figure3}b, where the continuous line is the result of a fit).

\unskip

\begin{figure}[h!]
	\setlength{\unitlength}{\textwidth}
	\begin{picture}(0.65,0.65)
		\put(0.175,0.)
		{
			\includegraphics[width=0.65\unitlength]{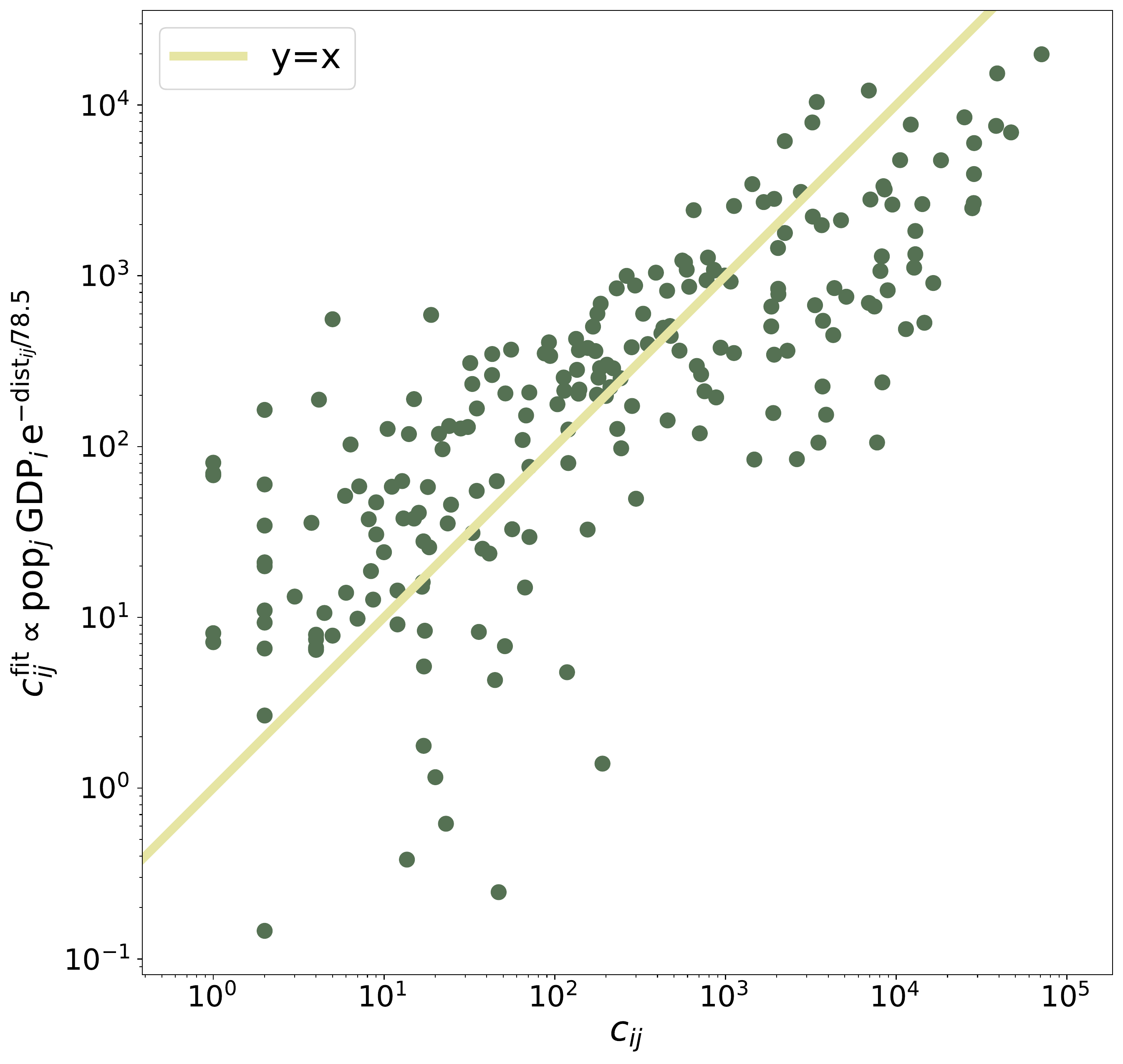}
		}
				
	\end{picture}
	\caption{Actual commuters' flux \textit{vs} the flux reconstructed by a fitted model that decays exponentially with the distance, and grows linearly with the GDP of the region of destination. The continuous line is the identity line. Only non-zero entries of the commuters' matrix are displayed and considered in the fitting procedure.}
	\label{figure4}
\end{figure}

The two determinants (distance and GDP) are considered together in Figure~\ref{figure4}, where the actual number of commuters (from one region to another; the flux is not symmetric) is compared to the result of the fitted model; the good agreement of the reconstructed flux with the real one (the continuous line is the identity line) supports the assumptions of the model.

\unskip

Starting from these results, we make the hypothesis that the total flux of people between regions is made of the commuters flux plus an `episodic' flux, unknown but with the same functional form (exponential decay with distance; linear dependence on GDP). Then, we built, on the total-flux matrix, a SIR-like model \cite{hethcote2000mathematics} that takes into account the exchange of infected people between regions. Temperature impacts the model in two ways. The first increases with the temperature, and is akin to the high-temperature (rightmost) arm of the model of Figure~\ref{figure2}a. The other modulates the contagiousness of the disease (higher for lower temperatures).

In the model, each month a new disease starts spreading from a given region (chosen according to a probability distribution); at end of the month, a fraction of the `recovered' people dies. Added to this effect are the high-temperature mortality, and finally a generic, temperature-independent, region-specific mortality.

\begin{figure}[h!]
	\setlength{\unitlength}{\textwidth}
	\begin{picture}(1,0.53)
		\put(0.,0.033)
		{
			\includegraphics[width=0.532\unitlength]{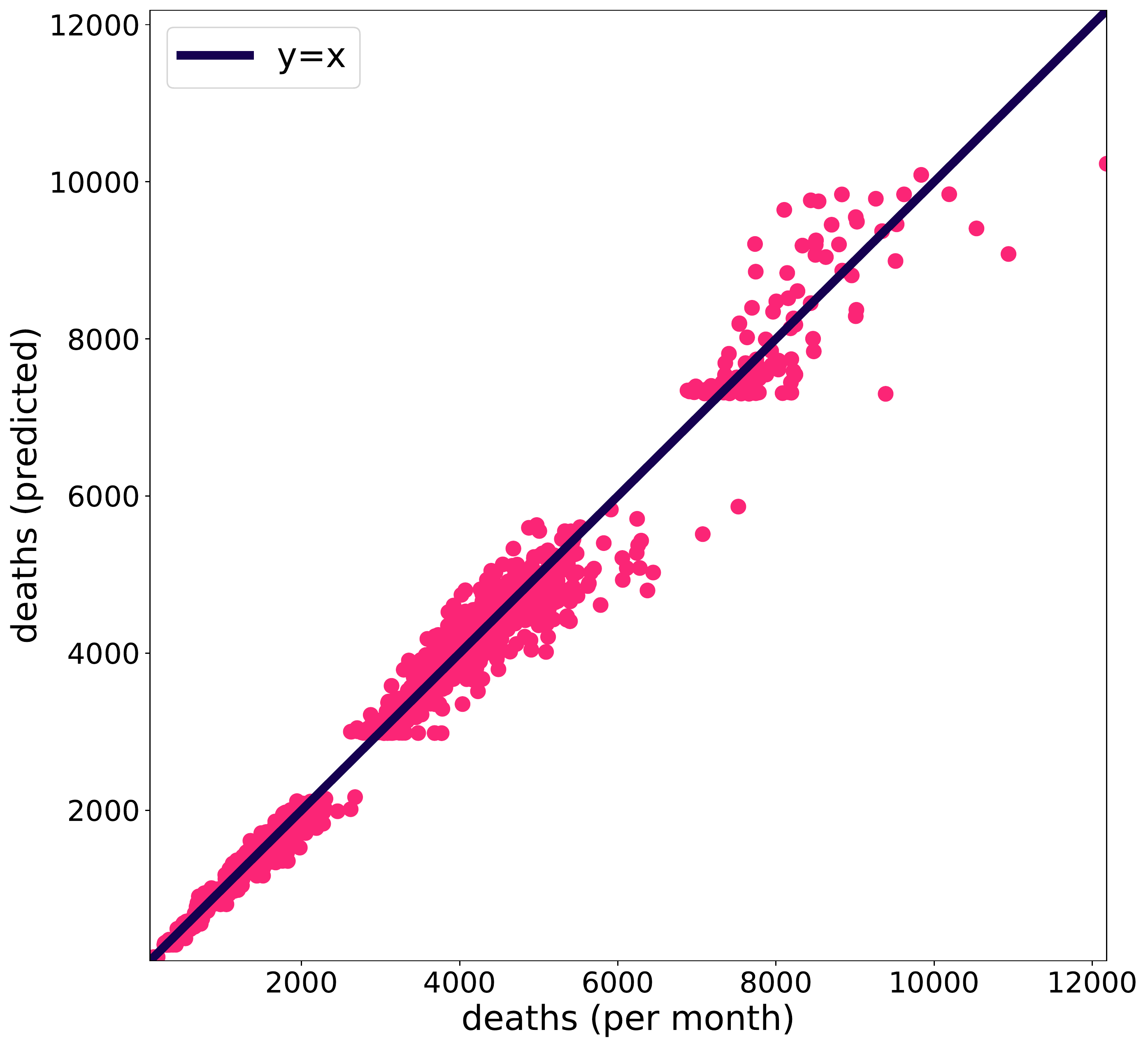}
		}
					   
		\put(0.52,0.025)
		{
			\includegraphics[width=0.49\unitlength]{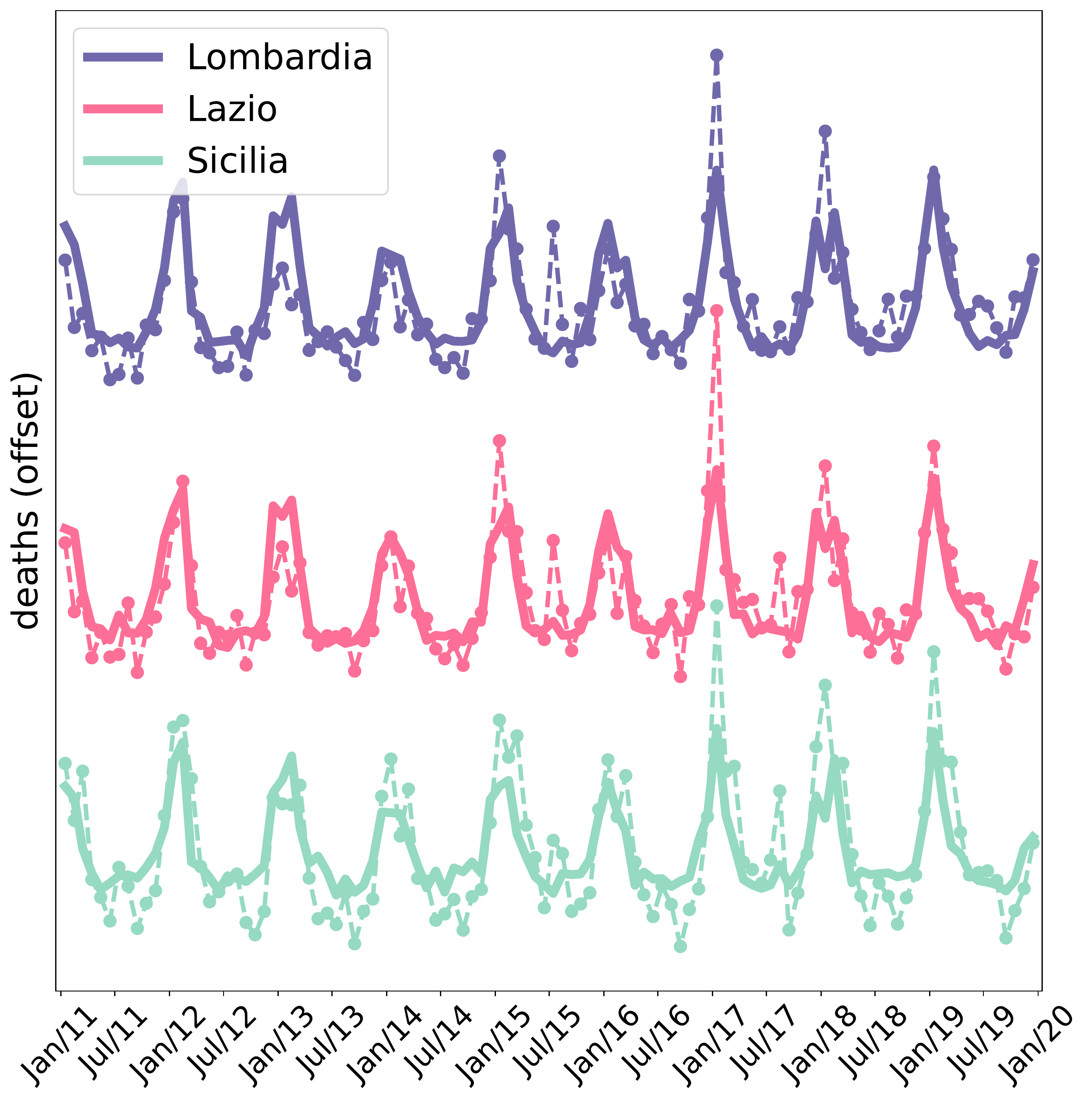}
		}
					   
		\put(0.25,0.){\makebox(0,0)[b]{(a)}}
		\put(0.75,0.){\makebox(0,0)[b]{(b)}}
					   
	\end{picture}
	\caption{\textbf{a}) Actual deaths \textit{vs} the deaths expected by the model (corr = 0.993; the continuous line is the identity). \textbf{b} Time-series of the deaths for three regions; dashed lines: data; continuous lines: model.}
	\label{figure5}
\end{figure}

We fitted the model's parameters (see Material and Methods) on the data. Figure~\ref{figure5}a shows the death counts for all the regions and all the considered months against the death counts generated by the model. The model can reproduce a large part of the observed variability (corr = 0.993; the continuous line is the identity line). This can be appreciated, as the deaths evolve in time, in Figure~\ref{figure5}b, for three different regions (dashed lines: data; continuous line: model). Note that the three time series are offset vertically such as to make the comparison data-\textit{vs}-model clearer.

\unskip

\begin{figure}[h!]
	\setlength{\unitlength}{\textwidth}
	\begin{picture}(0.65,0.65)
		\put(0.175,0.)
		{
			\includegraphics[width=0.65\unitlength]{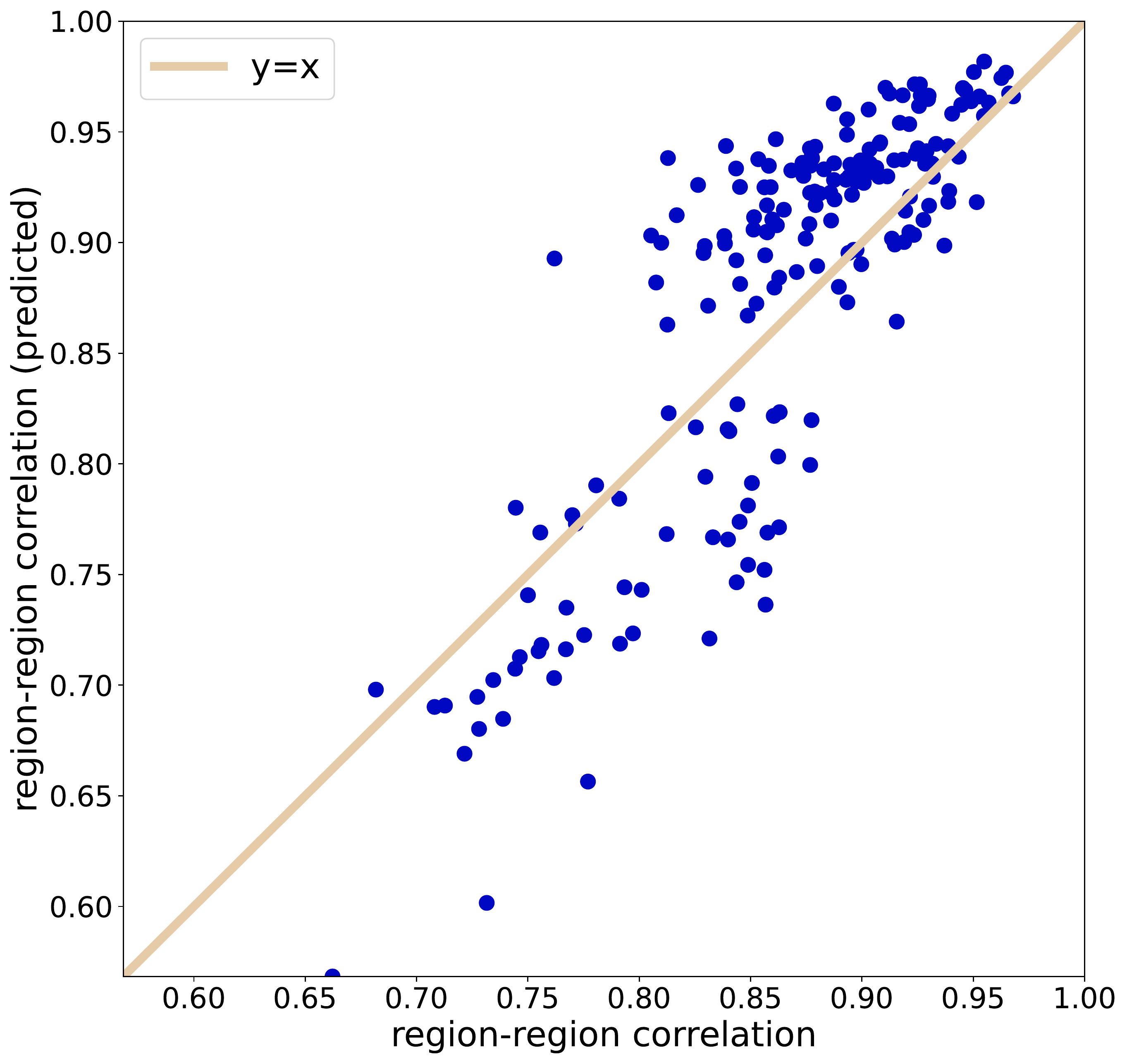}
		}
				
	\end{picture}
	\caption{Between-region correlation: data \textit{vs} reconstructed from the model (corr = 0.841; the continuous line is the identity line).}
	\label{figure6}
\end{figure}

Finally, we compare, in Figure~\ref{figure6}, the observed between-region correlations and the correlations between the time series produced by the model for each region. To a large extent, the model can capture the variability of the correlation among the regions (corr = 0.841; the continuous line is the identity line).

\unskip

\section{Discussion}
We aptly reconstructed the strong correlation among temporal series of all-cause monthly death rates relative to the 20 Italian regions by a model encompassing non-infectious (mainly summer) and infectious (winter) components. This last component was modelled in terms of a set of SIR equations taking into account both the across regions commuters exchange (daily flow) and the more irregular travellers' flux (longer time flow). The different mathematical treatments of summer and winter components allowed for a neat increase in the reconstruction of both global death rates and among-regions correlation strength concerning crude seasonality.

The reconstruction of the observed correlation network by our model (Figure~\ref{figure6}), excluding the hypothesis of contemporary arising `epidemic sources' across all the regions (that has near-zero probability), confirms the reliability of the proposed model. Overall, we can consider Italy as a proper `integrated system' that, thanks to both a rich exchange flux among regions and the sharing of `heat waves', reaches a general coherence in death rate fluctuations. This coherent behaviour acts as a largely invariant `mean field' governing all-cause monthly (and thus unaffected by longer time fluctuations of age class distribution) death rate fluctuations.

The existence of a very stable correlation network among Italian regions can be profitably used as a tool for the epidemiological surveillance of the territory: the arising of an anomalous value of the correlation degree of a region can be intended as the presence of an emerging source of risk (of both infectious and/or environmental origin). Thanks to the intrinsic redundancy of the correlation matrix, any (even transient) change reverberates on the entire network so allowing for a more sensible detection of instabilities: one clear example is the case of Recurrence Quantification Analysis (RQA, \cite{marwan2007recurrence}). RQA relies upon the construction of a distance (correlation in the case of angular metrics) matrix between subsequent epochs of a time (or space \cite{colafranceschi2005structure}) dependent signal. When in presence of regime changes driven by a slowly varying control parameter, RQA metrics (at odds with usual statistical indexes) exactly determine the entity and time (spatial) location of regime change \cite{trulla1996recurrence,casdagli1997recurrence}. Analogous considerations hold for correlation matrices and any other network system \cite{liu2022network} induced by external (\textit{e.g.} epidemics) or internal (\textit{e.g.} deterioration of living conditions) driving forces \cite{gorban2010correlations}.

The application of statistical-mechanics-inspired tools in public health is still in its infancy \cite{park2021network,huo2016dynamical} and/or confined to very specific issues \cite{sangeet2022quantifying,kitsak2010identification}. In this work, relying upon a very general consideration by Alexander Gorban and colleagues ``It is useful to analyse correlation graphs'' \cite{GORBAN202215}, we demonstrate how a raw (albeit very reliable) indicator, as all-cause mortality, is amenable to a statistical mechanics approach opening new avenues to epidemiological and environmental research.

\vspace{6pt} 

\section{Data sources}
The daily deaths for each Italian district (``comune''), starting from 2011, can be downloaded from the site of the Italian Institute of Statistics (ISTAT), following this link: \url{https://www.istat.it/storage/dati_mortalita/Dataset-decessi-comunali-giornalieri_regioni\%28excel\%29_5-21-ottobre-2021.zip}.
	
The geographical coordinates for each \textit{comune} can be found in the following GitHub project: \url{https://github.com/MatteoHenryChinaski/Comuni-Italiani-2018-Sql-Json-excel}, notably the file italy\_geo.xlsx.
	
The data concerning the temperature have been obtained, on June 23th 2022, from the National Centers for Environmental Information (\url{https://www.noaa.gov/}), with the following order specifications:
\begin{itemize}
	\item	Begin date: 2012-01-01 00:00;
	\item	End date: 2021-12-31 23:59;
	\item	Data types: PRCP, SNWD, TAVG, TMAX, TMIN;
	\item   Custom Flags: Station Name, Geographic Location, Include Data Flags.
\end{itemize}
	
The number of daily commuters between regions has been obtained from the data made available by ISTAT at the link: \url{https://www.istat.it/storage/cartografia/matrici_pendolarismo/matrici_pendolarismo_2011.zip} (data description can be found at \url{https://www.istat.it/it/archivio/157423}). The number of commuters we used is the sum of commuters from any \textit{comune} belonging to one region towards any \textit{comune} of another region.
	
The GDP for each region has been retrieved from the Wikipedia page \url{https://it.wikipedia.org/wiki/Regioni_d\%27Italia} (GDP is ``Prodotto interno lordo'' or PIL, in Italian) on October 10th 2022.

\bibliographystyle{ieeetr}
\bibliography{references}

\end{document}